\documentclass[12pt,letterpaper]{article}        
\usepackage{jheppub}
\usepackage{epsfig}
\usepackage{amssymb,amsfonts}
\usepackage{hyperref}

\newcommand{\beq}{\begin{equation}}
\newcommand{\eeq}{\end{equation}}
\def\bea{\begin{eqnarray}}
\def\eea{\end{eqnarray}}
\newcommand{\nn}{\nonumber \\}

\title{\large Universal low temperature theory of\\ charged black holes with AdS$_2$ horizons}

\author{Subir Sachdev}

\affiliation{Department of Physics, Harvard University, Cambridge MA 02138 USA}
\affiliation{Perimeter Institute for Theoretical Physics, Waterloo, Ontario N2L 2Y5, Canada}

\emailAdd{sachdev@g.harvard.edu}

\abstract{We consider the low temperature quantum theory of a charged black hole of zero temperature horizon radius $R_h$, in a spacetime which is asymptotically AdS$_{D}$ ($D > 3$) far from the horizon. At temperatures $T \ll 1/R_h$, the near-horizon geometry is AdS$_2$, and the black hole is described by a universal 0+1 dimensional effective quantum theory of time diffeomorphisms with a Schwarzian action, and a phase mode conjugate to the U(1) charge. We obtain this universal 0+1 dimensional effective theory starting from the full $D$-dimensional Einstein-Maxwell theory, while keeping quantitative track of the couplings. The couplings of the effective theory are found to be in agreement with those expected from the thermodynamics of the $D$-dimensional black hole.}

\begin{document}

\maketitle

\pdfoutput=1
\pagestyle{plain} \setcounter{page}{1}
\newcounter{bean}
\baselineskip16pt

\section{Introduction}

\begin{figure}[tb]
\begin{center}
\includegraphics[width=6in]{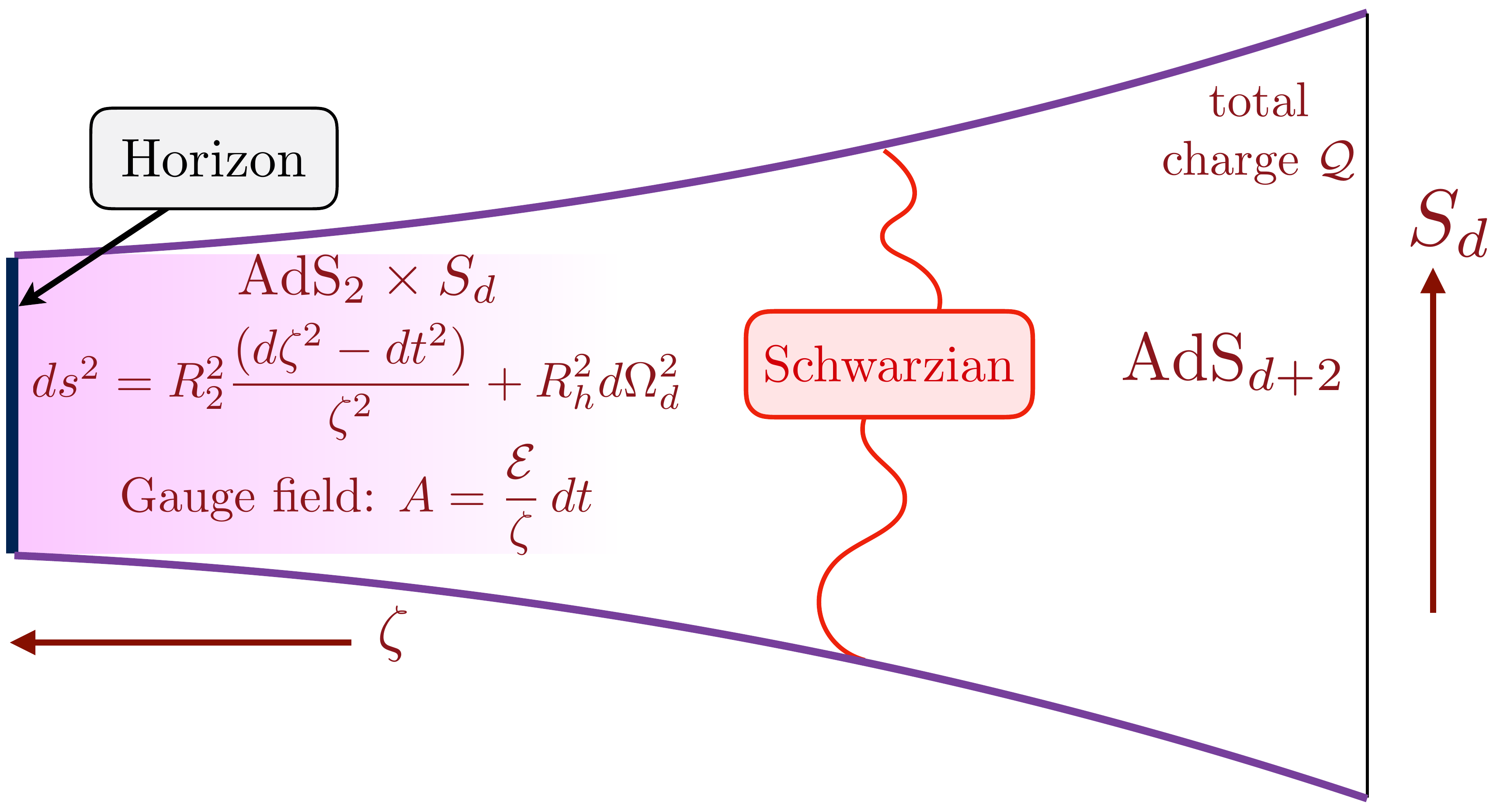}
\caption{Charged black holes in asymptotically AdS$_{d+2}$ space with a near-horizon AdS$_2 \times S_d$ geometry. 
We choose the compact space $M_d$ to be $S_d$, the $d$-sphere.
The intermediate geometry is described by the effective action in Eq.~(\ref{Seff}).} 
\label{fig:geometry}
\end{center}
\end{figure}

Charged black holes in asymptotically AdS$_{D}$ spacetimes 
have a near-horizon geometry, AdS$_2 \times M_{d}$, where $M_{d}$ is a compact space, and $D \equiv d+2$; see Fig.~\ref{fig:geometry}. The presence of the AdS$_2$ factor implies universal low energy quantum theories for such black holes. At sufficiently low energy scales, non-constant modes on $M_{d}$ are not excited, and much has been learnt about the resulting theories whose form depends only upon the conserved U(1) charges and the supersymmetry  \cite{Sen05,Dabholkar:2004yr,Dabholkar:2006tb,Sen08,SS10,Dabholkar:2014ema,AAJP15,Alm2016fws,kitaev2015talk,SS15,JMDS16,JMDS16b,HV16,KJ16,Fu:2016vas,Stanford:2017thb,Davison17,Gaikwad:2018dfc,Nayak:2018qej,Moitra:2018jqs,Chaturvedi:2018uov,Susskindtalk}. UV complete theories which realize these low energy limits are found in complex Sachdev-Ye-Kitaev models \cite{SY92,GPS01,SS15}, and they are also expected to appear in the low energy limit of supersymmetric string theories.

A common property of these black holes with charge $\mathcal{Q}$ is that their entropy $S(\mathcal{Q},T)$ at low $T$ has the form
\beq
S(\mathcal{Q},T \rightarrow 0) = S_0 (\mathcal{Q})  + \gamma \, T + \ldots\,, \label{defgamma}
\eeq
where the zero temperature limit $S_0 (\mathcal{Q})$ is non-zero. The recent advances concern the linear-in-$T$ term with co-efficient $\gamma$, which is determined by corrections to the purely AdS$_2$ near-horizon geometry. It has recently been recognized that these corrections are also universal \cite{AAJP15,Alm2016fws,JMDS16b}, and described by a Schwarzian effective action \cite{kitaev2015talk}.

For charged black holes, it is also important to consider the variation in the entropy as a function of $\mathcal{Q}$. In particular, an important dimensionless parameter is $\mathcal{E}$, defined by
\beq
\frac{dS_0 (\mathcal{Q})}{d\mathcal{Q}} = 2 \pi \mathcal{E} \quad, \quad T=0.\label{defE}
\eeq
The electric field in the near horizon region of the black hole is also determined by $\mathcal{E}$ \cite{Sen05}, as we shall see in Eq.~(\ref{Azeta}). The relationship in Eq.~(\ref{defE}) appeared in the context of complex SYK models \cite{GPS01}, before also appearing in the black hole context \cite{Sen05}, as was recognized later \cite{SS15}. The parameter $\mathcal{E}$ also determines the particle-hole asymmetry of probe matter fields in the AdS$_2$ region: {\it e.g.\/} a fermion of unit charge and scaling dimension $\Delta$ has the Green's function \cite{Faulkner09,Faulkner:2011tm}
\beq
G(\tau) \sim \left\{ 
\begin{array}{ccc}
-\tau^{-2\Delta}\, &~& \tau > 0 \\
e^{- 2 \pi \mathcal{E}} (-\tau)^{-2 \Delta} &~& \tau < 0 
\end{array}
\right. \quad, \quad T=0\,; \label{GE}
\eeq
this form applies also to the complex SYK models.

The universal low temperature quantum theory describes both energy and charge fluctuations. It is expressed in terms of a 
monotonic time diffeomorphism $f(\tau)$ obeying 
\beq
f(\tau + 1/T) = f(\tau) + 1/T \,, 
\eeq
and a phase phase field $\phi (\tau)$ obeying 
\beq
\phi (\tau + 1/T) = \phi (\tau) + 2 \pi n\,, 
\eeq
$n$ integer, which is conjugate to the total integer charge $\mathcal{Q}$. In the absence of 
supersymmetry, symmetry arguments lead to the following imaginary time action in the grand canonical ensemble \cite{Davison17}
\bea
I_{\rm eff} [f, \phi] &=& - S_0 (\mathcal{Q}) +  \frac{K}{2} \int_0^{1/T} d \tau (\partial_\tau \phi - i (2 \pi \mathcal{E} T) \partial_\tau f)^2 \nn &~&~~~~~~~~~ -  \frac{\gamma}{4 \pi^2} \int_0^{1/T} d \tau \, \{ \tan (\pi T f(\tau)), \tau\}, \label{Seff}
\eea
where we have introduced the Schwarzian
\beq
 \{ g(\tau), \tau\} \equiv \frac{g'''}{g'} - \frac{3}{2} \left( \frac{g''}{g'} \right)^2 \,.
 \eeq
This action is characterized by three parameters, $\gamma$, $K$, and $\mathcal{E}$, and these can be specified by their connection to thermodynamics. These parameters depend upon the charge $\mathcal{Q}$ (or the chemical potential $\mu$), but this dependence has been left implicit. 
We have already described the connections of $\gamma$ and $\mathcal{E}$ to the thermodynamics above. The parameter $K$ is the zero temperature compressibility
\beq
K = \frac{d \mathcal{Q}}{d \mu} \quad, \quad T=0 \,.
\eeq
A different 0+1 dimensional super-Schwarzian action is expected for supersymmetric black holes \cite{Fu:2016vas,Stanford:2017thb}, 
which we will not discuss here.

For neutral black holes connected to the Majorana SYK theory, the 0+1 dimensional theory has allowed non-perturbative computation of the density of low-energy states \cite{Cotler:2016fpe,Garcia-Garcia:2017pzl,Bagrets:2017pwq,Stanford:2017thb,Kitaev:2018wpr}. The phase action in Eq.~(\ref{Seff}) can extend such computations to charged black holes, and this is described in other recent papers \cite{GKST19,Liu:2019niv}. Contrary to early speculations \cite{Jensen:2011su}, the zero temperature entropy, $S_0 (\mathcal{Q})$, is not associated with an exponentially large degeneracy of the ground state (except in cases with $\mathcal{N}=2$ supersymmetry \cite{Sen05,Dabholkar:2004yr,Sen08,Dabholkar:2014ema,Fu:2016vas}). Instead, there is an exponentially small level spacing down to the ground state, and the envelope of the resulting density of states can be computed from $I_{\rm eff}$.

This paper will start from the Einstein-Maxwell 
theory of spherical black holes in asymptotically AdS$_{d+2}$ space, which we review in Section~\ref{sec:bh}. At low temperatures, such black holes are dominated by fluctuations in the near-horizon AdS$_2$ geometry \cite{Faulkner09}, and this is reviewed in our notation in Section~\ref{sec:ads2}. Section~\ref{sec:2dgravity} describes a further dimensional reduction from AdS$_2$ to the 0+1 dimensional Schwarzian theory, while keeping quantitative track of all couplings from the parent AdS$_{d+2}$ theory. This analysis yields the precise co-efficient, $\gamma$, of the Schwarzian action in terms of the couplings in the Einstein-Maxwell theory. This value of $\gamma$ is found to be just that expected from a match between the thermodynamics of the black hole in AdS$_{d+2}$ and the Schwarzian theory. 

We turn our attention to the action for the phase mode, $\phi$, in Eq.~(\ref{Seff}) in Section~\ref{sec:phase}. Here, we identify $\phi$ with the value of a Wilson line extending from the black hole horizon to the AdS$_{d+2}$ boundary: see Eq.~(\ref{defphi}), and also Refs.~\cite{Nickel:2010pr,Moitra:2018jqs}. We present arguments, based largely on gauge invariance, which lead to a derivation of the $K$ term in Eq.~(\ref{Seff}).

\section{Black holes in asymptotically AdS space}
\label{sec:bh}

We consider the case of spherical black holes in global AdS$_{d+2}$ ($d > 1$), following the analysis of Chamblin {\it et al.\/} \cite{Myers99}.
The Einstein-Maxwell theory of a metric $g$ and a U(1) gauge flux $F = dA$ has Euclidean action
\beq
I_{EM} =  \int d^{d+2} x \sqrt{g} \left[ -\frac{1}{2 \kappa^2} \left(\mathcal{R}_{d+2} + \frac{d(d+1)}{L^2} \right) +  \frac{1}{4g_F^2} F^2 \right], \label{EM}
\eeq
where $\kappa^2 = 8 \pi G_N$ is the gravitational constant, $\mathcal{R}_{d+2}$ is the Ricci scalar, 
$L$ is the radius of AdS$_{d+2}$, and $g_F$ is a U(1) gauge coupling constant.
We will not assume any particular value for the length ratio $L/R_h$, and only assume that it is kept fixed as we take the $T \rightarrow 0$ limit.
We choose a solution of the saddle-point equations of Eq.~(\ref{EM}) with metric
\beq
ds^2 =  V(r) d\tau^2 + r^2 d \Omega_d^2 + \frac{dr^2}{V(r)} \label{s1}
\eeq
where $d \Omega_d^2$ is the metric of the $d$-sphere, and
\beq
V(r) = 1 + \frac{r^2}{L^2} + \frac{\Theta^2}{r^{2d-2}} - \frac{M}{r^{d-1}}.
\eeq
Note that as $r \rightarrow \infty$, the metric in Eq.~(\ref{s1}) is AdS$_{d+2}$ with boundary geometry $S_d \times S_1$;
here $S_d$ is a sphere with a $d$-dimensional surface, and $S_1$ represents the thermal circle.

The gauge field solution has the form
\beq
A = i \mu \left( 1 - \frac{r_0^{d-1}}{r^{d-1}} \right) d \tau \label{a1}
\eeq
The equations (\ref{s1}) and (\ref{a1}) solve the Einstein-Maxwell equations provided
\beq
\Theta = \sqrt{\frac{(d-1)}{d}} \frac{\kappa r_0^{d-1}}{g_F} \mu  \label{a2}
\eeq
The value of the gauge field at the AdS boundary defines the chemical potential $\mu$, provided $r_0$ is the horizon. This in turn demands that $V(r_0)=0$ or
\beq
M = r_0^{d-1} \left( 1 + \frac{r_0^2}{L^2} + \frac{\Theta^2}{r_0^{2d-2}} \right). \label{a3}
\eeq
The temperature of the black hole, $T$, is given by
\beq
4 \pi T = V'(r_0)\,. \label{a4}
\eeq
The Eqs.~(\ref{a2},\ref{a3},\ref{a4}) determine all the parameters, $\Theta$, $M$, $r_0$ in terms of $\mu$ and $T$. So we have determined a unique black hole solution in terms of the independent thermodynamic parameters $\mu$ and $T$.

Let us now specify the thermodynamic potentials of the black hole solution. 
We can compute the grand potential, $\Omega (\mu, T)$, by evaluating the action in Eq.~(\ref{EM}) for the solution above. 
However, to obtain a finite answer as the boundary of spacetime at $r=r_\infty \rightarrow \infty$, we have to include boundary 
counterterms to render the action finite. One of these terms is the familiar Gibbons-Hawking term:
\beq 
I_{GH} = -\frac{1}{\kappa^2} \int_{\partial} d^{d+1} x \, \sqrt{g_b} \,  \mathcal{K}_{d+1} \label{IGH}
\eeq
where the boundary has induced metric $g_b$, and trace of the extrinsic curvature $\mathcal{K}_{d+1}$. In addition, the CFT$_{d+1}$ residing on the boundary requires local counterterms to obtain a finite action, and these are \cite{Henningson:1998ey,Balasubramanian:1999re,Myers99a,Myers99,Nayak:2018qej}
\beq
I_{ct} =  \frac{1}{\kappa^2} \int_{\partial} d^{d+1} x \sqrt{g_b} \left[ \frac{d}{L} + \frac{L}{2 (d-1)} \mathcal{R}_{d+1} + \ldots \right]\, \label{Ict}
\eeq
where the boundary has Ricci scalar $\mathcal{R}_{d+1}$, and we have only shown terms that are needed in $d=2$. 
We list the individual contributions of the different actions:
\bea
T I_{EM} &=& \frac{s_d}{\kappa^2} \left( - \frac{(d-1) }{2 L^2}(r_\infty^{d+1} - r_0^{d+1}) - \frac{(d+1)}{2} (r_\infty^{d-1} - r_0^{d-1})\right) + \frac{s_d (d-1)^2 r_0 \mu^2}{2dg_F^2} \nn
T I_{GH} &=& \frac{s_d}{\kappa^2} \left( - \frac{(d+1)r_\infty^{d+1}}{2 L^2} - \frac{(d-1)r_\infty^{d-1}}{2} \right) \nn
T I_{ct} &=& \frac{s_d}{\kappa^2} \left(  \frac{2 r_\infty^3}{L^2} + 2 r_\infty - r_0 - \frac{r_0^3}{L^2} \right) 
- \frac{2 \pi r_0 \mu^2}{g_F^2} \quad , \quad d=2\,,
\eea
where $s_d \equiv 2 \pi^{(d+1)/2} \Gamma((d+1)/2)$ is the area of $S_d$ with unit radius.
Combining the actions, the terms diverging as $r_\infty \rightarrow \infty$ cancel, and 
the grand potential is \cite{Myers99}
\bea
\Omega (T, \mu) &=& T(I_{EM} + I_{GH} + I_{ct}) \nn
&=& \frac{s_d [r_0 (T, \mu)]^{d-1}}{2 \kappa^2} \left(1 - \frac{[r_0 (T, \mu)]^2}{L^2} \right) - \frac{s_d (d-1) \mu^2 [r_0 (T, \mu)]^{d-1}}{2d g_F^2}\,. \label{Omega}
\eea
We can now evaluate the entropy by taking the temperature derivative of $\Omega$ to obtain 
\beq
S(T, \mu) = \frac{2 \pi s_d}{\kappa^2} \, [r_0 (T,\mu)]^d, \label{s5}
\eeq
which is precisely the expression expected from Hawking's formula.
Similarly, the total charge is obtained by taking the $\mu$ derivative of $\Omega$
\beq
\mathcal{Q}(T, \mu) =\frac{s_d (d-1) \mu \, [r_0(T,\mu)]^{d-1}}{ g_F^2} \,, \label{s6}
\eeq
and this expression can also be obtained from Gauss's law evaluated as $r \rightarrow \infty$

All results above apply for general $T$ and $\mu$, and the $T$ and $\mu$ dependence of $r_0$ can be obtained from Eqs.~(\ref{a2},\ref{a3},\ref{a4}).
Let as us now turn to a consideration of the low $T$ limit. 
Explicitly, we have for $r_0$
\beq
r_0 (T, \mu) = R_h   + \frac{2 \pi L^2}{d+1} T + \mathcal{O}(T^2) \quad, \quad \mbox{$T \rightarrow 0$, $\mu$ fixed}\,,
\eeq
where $R_h$ is the radius the black hole horizon at $T=0$
\beq
R_h \equiv \frac{L}{g_F} \left[\frac{(d-1)( \mu_0^2 \kappa^2 (d-1) - d g_F^2)}{d(d+1)} \right]^{1/2}\,, \label{rh}
\eeq
with $\mu_0 \equiv \mu (T=0)$. 
Note that the size of the black hole at $T=0$ is determined by the chemical potential $\mu_0$ alone, and $\mu_0$ has to be large enough so that the expression inside the square root is positive. We can invert Eq.~(\ref{rh}) to write
\beq
\mu_0 =  \frac{g_F}{L \kappa (d-1)} \left[ d \left( (d+1) R_h^2 + (d-1) L^2 \right) \right]^{1/2}\label{mu0}
\eeq
For the charge $\mathcal{Q}$ in Eq.~(\ref{s6}) we have
\beq
\mathcal{Q} = -\left( \frac{\partial \Omega}{\partial \mu} \right)_{T} = \frac{s_d R_h^{d-1} \sqrt{d \left[ (d+1)R_h^2  + (d-1)L^2  \right]}}{L \kappa g_F} \quad , \quad T=0 \,,
\label{Q1}
\eeq
Below we will express all the low $T$ thermodynamic parameters of the black hole in terms of $R_h$.
These results can be converted to a dependence on $\mathcal{Q}$ or $\mu_0$ via Eqs.~(\ref{Q1}) or (\ref{mu0}).

For the grand potential at $T=0$ we obtain
\beq
\Omega_0 = - \frac{d R_h^2 s_d}{(d-1) L^2 \kappa^2} \quad , \quad T=0\,,
\eeq
while the $T=0$ entropy in Eq.~(\ref{s5}) is
\beq
S_0 =  -\left( \frac{\partial \Omega}{\partial T} \right)_{\mu} = \frac{2 \pi s_d}{\kappa^2} \, R_h^d \quad, \quad T=0\,. \label{S1}
\eeq
We can obtain the function $S_0 (\mathcal{Q})$ by eliminating $R_h$ between Eqs.~(\ref{Q1}) and (\ref{S1}). Also we can compute the compressibility
\beq
K = \left. \frac{d \mathcal{Q}}{d \mu} \right|_{T=0} =  \frac{d \mathcal{Q}/dR_h}{d \mu_0/dR_h} = \frac{(d-1) s_d R_h^{d-3} \left[ d(d+1) R_h^2
+ (d-1)^2 L^2 \right]}{(d+1) g_F^2}\,. \label{defK}
\eeq
We also quote another derivative we will need later in Section~\ref{sec:phaseT}
\beq
\left( \frac{\partial^2 \Omega}{\partial T^2} \right)_{\mu} =
-\frac{4 d \pi^2 s_d L^2 R_h^{d-1}}{(d+1)
   \kappa ^2} \quad , \quad T=0 \,. \label{DeltaOmega0}
\eeq

For the analysis of the low $T$ limit, it is better to work at fixed $\mathcal{Q}$ rather than fixed $\mu$. In the SYK model, the intermediate frequency structure at $T \ll \omega \ll J$ remains independent of $T$ only when we work at fixed $\mu$ \cite{PGKS97,GPS01,SS15}. We will find a similar feature in the low $T$ limit of the present black hole solution below. So as $T \rightarrow 0$, we write
\beq
\mu = \mu_0 - 2 \pi \mathcal{E} T + \ldots \quad, \quad \mbox{$T \rightarrow 0$, $\mathcal{Q}$ fixed} \label{muET}
\eeq
where 
\beq 
2 \pi \mathcal{E} \equiv - \left( \frac{\partial \mu}{\partial T} \right)_{\mathcal{Q}} = \left(\frac{\partial S}{\partial \mathcal{Q}} \right)_T \,. \label{EmuS}
\eeq
The first equality in Eq.~(\ref{EmuS}) is the definition of $\mathcal{E}$ which follows from the expansion in Eq.~(\ref{muET}), while the second equality is a general thermodynamic Maxwell relation. We will see below in Eq.~(\ref{Azeta}) that $\mathcal{E}$ also specifies the electric field at the surface of the black hole.
We can compute the value of $\mathcal{E}$ from the definition in Eq.~(\ref{EmuS}) and Eqs.~(\ref{a2},\ref{a3},\ref{a4},\ref{s6}), and obtain
\beq
\mathcal{E} = \frac{g_F R_h L \sqrt{d \left[ (d+1) R_h^2 + (d-1)L^2 \right]}}{\kappa \left[ d(d+1)R_h^2+  (d-1)^2 L^2 \right]}. \label{s4}
\eeq
The `equation  of state' obeyed by $\mathcal{E}$ and $\mathcal{Q}$ is obtained by eliminating $R_h$ between 
Eqs.~(\ref{Q1}) and (\ref{s4}); this leads to a lengthy expression which we shall not write out explicitly. But, 
we can use Eqs.~(\ref{Q1}), (\ref{S1}) and (\ref{s4}) to verify the identity in Eq.~(\ref{defE}) 
\beq
\frac{dS_0}{d\mathcal{Q}} = \frac{\partial S_0/\partial R_h}{\partial \mathcal{Q}/\partial R_h} = 2 \pi \mathcal{E} \quad, \quad T=0.\label{s66}
\eeq

Using Eq.~(\ref{muET}), we can compute the variation of the entropy at fixed $\mathcal{Q}$, and so obtain
\beq
\gamma = \left( \frac{\partial S}{\partial T} \right)_Q = 
\frac{4 \pi^2 d s_d L^2 R_h^{d+1}}{\kappa^2 (d(d+1) R_h^2 + (d-1)^2 L^2)}\,. \label{gammares1}
\eeq
We will match this to the co-efficient of the Schwarzian below in Eqs. (\ref{gammares}) and (\ref{Seff1}).

\subsection{Dimensional reduction}
\label{sec:ads2}

\begin{figure}[tb]
\begin{center}
\includegraphics[width=5in]{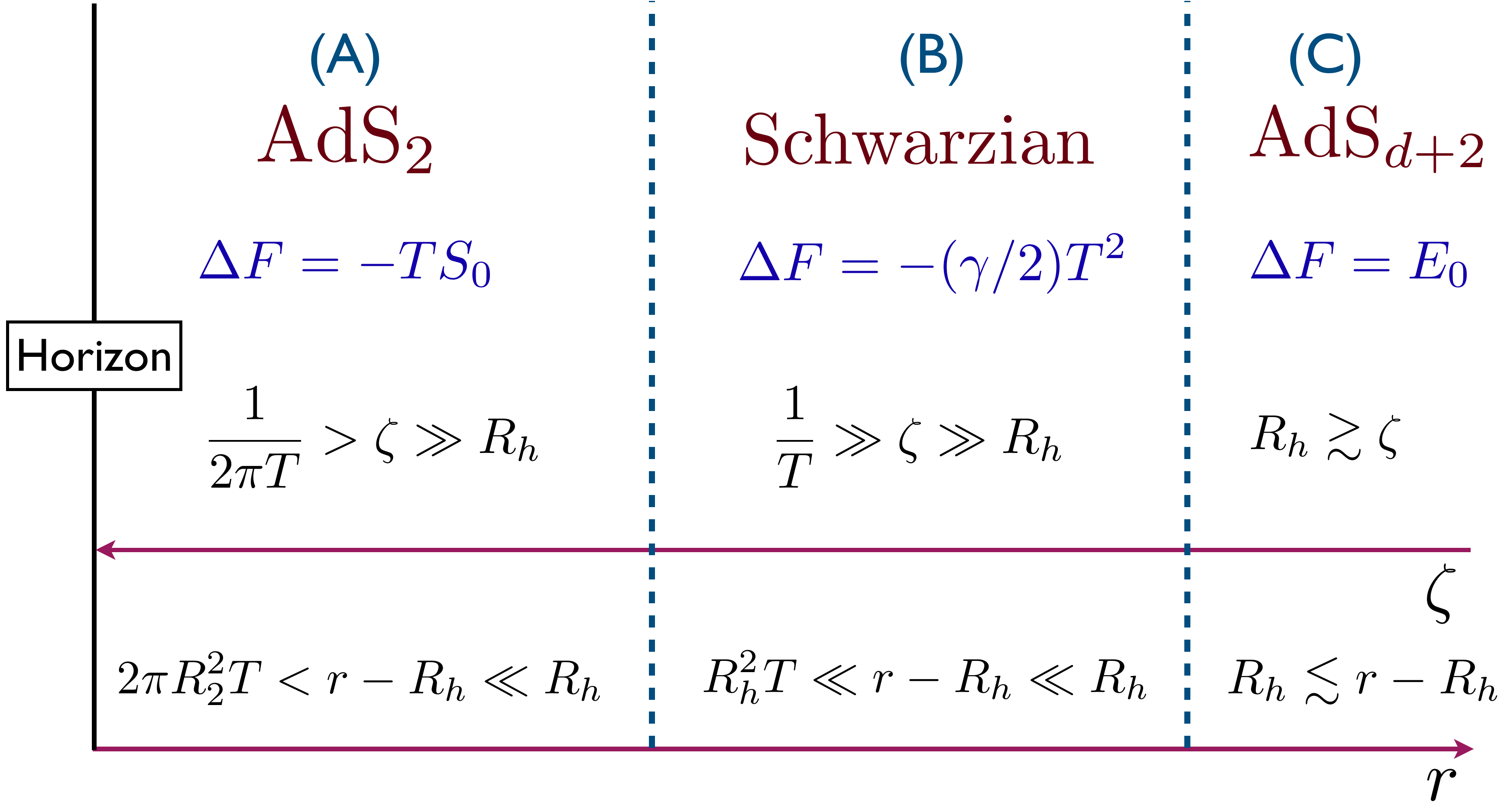}
\caption{Low temperature crossovers outside a black hole of charge $\mathcal{Q}$. The value of $R_h$ is determined from $\mathcal{Q}$ via Eq.~(\ref{Q1}), and we describe $T \ll 1/R_h$ at fixed $\mathcal{Q}$, $\mu$ specified by Eq.~(\ref{muET}), and $R_2 \sim R_h$. We denote contributions to the free energy $F = \Omega + \mu \mathcal{Q}$ in each region ($E_0 = \Omega_0 + \mu_0 \mathcal{Q}$ is the ground state energy).} 
\label{fig:bh}
\end{center}
\end{figure}

As illustrated in Figs.~\ref{fig:geometry} and  \ref{fig:bh}, the black hole solution exhibits interesting crossovers in the near-horizon region, when 
\beq
T R_h \ll 1 \,. 
\eeq
Throughout we will assume that the AdS$_{d+2}$ radius, $L$ is held fixed as $T$ is lowered. So we take the $T \rightarrow 0$ limit at fixed, but arbitrary, $L/R_h$.

It is useful to introduce the co-ordinate $\zeta$ via
\beq
r  = R_h + \frac{R_2^2}{\zeta}, \label{rzeta}
\eeq
so that $T=0$ the horizon will be at $\zeta=\infty$. We choose the length scale $R_2$ to be
\beq
R_2  = \frac{L R_h}{\sqrt{d(d+1) R_h^2 + (d-1)^2 L^2}} \,, \label{valR2}
\eeq
and the reason for this specific choice for $R_2$ will become clear below. Note that as $T \rightarrow 0$, $R_2/R_h$ is fixed, 
and we assume in this subsection that we are in the 
near-horizon region defined by (see Fig.~\ref{fig:bh})
\beq
\zeta \gg R_h\,.
\eeq

Now we insert the co-ordinate change (\ref{rzeta}) in the metric (\ref{s1}), 
and expand in powers of $T R_h$ while assuming $\zeta \sim 1/T$. Before performing this expansion it is important that we fix the charge of the black hole at $\mathcal{Q}$ also at $T>0$ \cite{Faulkner09}. This requires that the chemical potential acquires the $T$-dependence in Eq.~(\ref{muET}).
With this $T$-dependent $\mu$, we find that the metric takes the form
\beq
ds^2 = \frac{R_2^2}{\zeta^2} \left[ (1 - 4 \pi^2 T^2 \zeta^2) d\tau^2 + \frac{d \zeta^2}{1 - 4 \pi^2 T^2 \zeta^2} \right] + R_h^2 d \Omega_d^2\,. \label{ads2}
\eeq
This metric is AdS$_2 \times S_d$ at $T=0$. But there is a co-ordinate transformation which maps the $T>0$ metric to the $T=0$ metric for AdS$_2$: we map the $(\tau,\zeta)$ co-ordinates to $(\tau_0, \zeta_0)$ co-ordinates via
\bea
 \tau_0 &=&  \frac{ (\pi T)^{-1} (1 - 4 \pi^2 T^2 \zeta^2)^{1/2} \, \sin(2 \pi T \tau)}{ 1 +  (1 - 4 \pi^2 T^2 \zeta^2)^{1/2} \, \cos(2 \pi T \tau)} \nn
\zeta_0 &=&  \frac{ 2 \zeta }{ 1 +  (1 - 4 \pi^2 T^2 \zeta^2)^{1/2} \, \cos(2 \pi T \tau)} \,. \label{map}
\eea
Then the metric for $(\tau_0, \zeta_0)$ is just as in Eq.~(\ref{ads2}) but with $T=0$. Also note that for small $\zeta$, the co-ordinate transformation becomes 
\beq
\tau_0 = g(\tau)\,, \quad \zeta_0 = \zeta g'(\tau)\,, \quad g(\tau) = \frac{\tan(\pi T \tau)}{\pi T}\,, \quad \zeta \rightarrow 0\,. \label{map2}
\eeq
We see from Eq.~(\ref{ads2}) that the horizon at non-zero $T \ll 1/R_h$ is at $\zeta = 1/(2 \pi T)$, and we are interested in the near-horizon region (A) in Fig.~\ref{fig:bh}
\beq
R_h \ll \zeta < \frac{1}{2 \pi T} \,. \label{nearhorizon}
\eeq
Note also the prefactor of $R_2^2$ in Eq.~(\ref{ads2}); the value of $R_2$ was chosen in Eq.~(\ref{valR2}) by anticipating this prefactor.

Turning to the gauge field sector, the solution in Eq.~(\ref{a1})
has the form 
\beq
A = i \frac{\mathcal{E}}{\zeta} (1 - 2 \pi T \zeta ) d \tau \,, \label{Azeta}
\eeq
where the dimensionless electric field $\mathcal{E}$ is exactly equal to that obtained from the definition in Eq.~(\ref{EmuS}). This can also be mapped onto the $T=0$ gauge field via the co-ordinate transformation in Eq.~(\ref{map}), but after a gauge transformation \cite{Faulkner:2011tm}.
Explicitly, let us write in the co-ordinate system at $T=0$, $A_{\tau_0} = \mathcal{E}/\zeta_0$. Then
\beq
A_{\tau_0} d \tau_0 = \overline{A}_\tau d \tau + \overline{A}_\zeta d \zeta\,,
\eeq
and $\overline{A}_{\tau,\zeta}$ can be computed from Eq.~(\ref{map}). We perform a gauge transformation generated by $\Lambda$ to obtain
$A_\tau = \overline{A}_\tau - \partial_\tau \Lambda$, $A_\zeta = \overline{A}_\zeta - \partial_\zeta \Lambda$. In this manner, we obtain $A_\zeta =0$,
and $A_\tau$ as in Eq.~(\ref{Azeta}), provided we choose
\beq
\Lambda = i 2 \pi T \mathcal{E} \tau - 2 i \mathcal{E} \tan^{-1} \left( \frac{1 - \sqrt{1 - 4 \pi^2 T^2 \zeta^2}}{\zeta} \tan( \pi T \tau) \right)\,. 
\eeq

It is useful to write out some of the thermodynamic parameters in terms of the parameters of the AdS$_2 \times$ S$_d$ geometry, but independent of $L$, the radius of AdS$_{d+2}$.
We can write the charge $\mathcal{Q}$ in Eq.~(\ref{Q1}) as
\beq
\mathcal{Q} = \frac{s_d}{g_F^2} \frac{ R_h^d}{R_2^2} \mathcal{E}\,, \label{QE}
\eeq
which is the expression expected from application of Gauss's law at the horizon using Eqs.~(\ref{ads2},\ref{Azeta}). And the linear-in-$T$ term in the entropy in Eq.~(\ref{gammares1}) can be written as
\beq
\gamma = \frac{4 \pi^2 d s_d R_2^2 R_h^{d-1}}{\kappa^2}\,. \label{gammares}
\eeq
Although there is no corresponding expression for $K$ which is independent of $L$, we do have the relation
\beq
\mu_0 = \frac{\mathcal{E} R_h}{(d-1)R_2^2 } \,. \label{mu0e}
\eeq
 The expression in Eq.~(\ref{mu0e}) relates the chemical potential to the work done by a unit charge moving from the horizon to the boundary.

We emphasize that the relations in Eqs.~(\ref{QE},\ref{gammares},\ref{mu0e}) hold at arbitrary values of the ratio $L/R_h$.

\section{The Schwarzian action}
\label{sec:2dgravity}

Section~\ref{sec:ads2} described the reduction of the spacetime metric from $d+2$ dimensions, a form which factorized spacetime into 2 and $d$ dimensions. Fluctuations in the $d$-dimensional space, which is a sphere of radius $\sim R_h \sim R_2$, are expected to be subdominant for $T \ll 1/R_h$. In this section we will perform the equivalent reduction in terms of the action, to an effective quantum gravity theory in 2 dimensions. Then we will follow Maldacena {\it et al.} \cite{JMDS16b} and further reduce the two-dimensional gravity to a one-dimensional Schwarzian action.

We write the $(d+2)$-dimensional metric $g$ of $I_{EM}$ in Eq.~(\ref{EM}) in terms of a two-dimensional metric $h$ and a scalar field $\Phi$ \cite{Davison17,Nayak:2018qej}:
\beq
ds^2 = \frac{ds_2^2}{\Phi^{d-1}} + \Phi^{2} \, d \Omega_d^2 \,. \label{dsds2}
\eeq
Both $h$ and $\Phi$, and the gauge field $A$, are allowed to be general functions of the two-dimensional co-ordinates $\zeta$ and $\tau$. To begin with, we do not specialize to just the near-horizon region, and so there are no restrictions on the value of $\zeta$ other than it is outside the horizon; from Eq.~(\ref{rzeta}) the latter constraint is $\zeta < R_2^2/(r_0 - R_h)$.
The boundary of the spacetime is at $\zeta \rightarrow 0$, corresponding to $r \rightarrow \infty$.
Then the expressions for the Einstein-Maxwell (Eq.~(\ref{EM})) and Gibbons-Hawking (Eq.~(\ref{IGH})) actions reduce to ($x \equiv (\tau, \zeta)$)
\bea
I_{EM} &=& \int d^2 x \sqrt{h} \left[ - \frac{s_d}{2\kappa^2} \, \Phi^d \, \mathcal{R}_2 + U(\Phi) + \frac{Z(\Phi)}{4 g_F^2} F^2 \right]  \nn
I_{GH} &=& - \frac{s_d}{\kappa^2} \int_{\partial} dx \sqrt{h_b} \Phi^d \, \mathcal{K}_1 \label{2d1}
\eea
along with an additional term not displayed which cancels in $I_{EM} + I_{GH}$ \cite{Nayak:2018qej}.
The Gibbons-Hawking term is to be evaluated at the boundary at $\zeta \rightarrow 0$ or $r \rightarrow \infty$.
Here $\mathcal{R}_2$ is the two-dimensional Ricci scalar, the second integral is over a one-dimensional boundary with metric $h_b$ and extrinsic curvature $\mathcal{K}_1$. 
The powers of $\Phi$ in Eq.~(\ref{dsds2}) were judiciously chosen so that there would be no gradient of $\Phi$ in Eq.~(\ref{2d1}), and that $\mathcal{R}_2$ would couple to $\Phi^d$.
The explicit forms of the potentials $U(\Phi)$ and $Z(\Phi)$ are,
\beq
U(\Phi) = - \frac{s_d}{2 \kappa^2} \left( \frac{d(d-1)}{\Phi} + \frac{d(d+1) \Phi}{L^2} \right) \quad, \quad Z(\Phi) = s_d \Phi^{2d-1}  \,. \label{2d2}
\eeq
The two-dimensional action in Eqs.~(\ref{2d1},\ref{2d2}) has exactly the same saddle point solution as that of the four-dimensional action in Eq.~(\ref{EM}), and this solution can be obtained by mapping Eq.~(\ref{dsds2}) to Eq.~(\ref{s1}). In particular, we obtain from this solution using Eqs.~(\ref{s1}) and (\ref{rzeta}) the exact expression for the saddle point value of $\Phi$
\beq
\Phi (\zeta) =  R_h + \frac{R_2^2}{\zeta} \,. \label{Phizeta}
\eeq
This scalar field profile will be a key ingredient in the derivation of the Schwarzian action below.

The next step is to renormalize the theory in Eq.~(\ref{2d2}) down to the near horizon region so that the spacetime is in the region defined by Eq.~(\ref{nearhorizon}), and the boundary of spacetime is at a $\zeta=\zeta_b$ in region (B) of Fig.~\ref{fig:bh}
\beq 
R_h \ll \zeta_b \ll \frac{1}{T} \,. \label{GHzeta}
\eeq
This moving of the boundary will induce a complicated renormalization of the potentials $V(\Phi)$ and $Z(\Phi)$, but it will turn out that we will not need the explicit form of this renormalization. The boundary term $I_{GH}$ in Eq.~(\ref{2d1}) will now be evaluated at $\zeta = \zeta_b$. Note that the counterterms, $I_{ct}$, in Eq.~(\ref{Ict}) all vanish when evaluated at a one-dimensional boundary with $d=0$ spatial dimensions: so they have no counterpart for two-dimensional gravity. Nevertheless, in computing the free energy of the theory in Eq.~(\ref{2d1}), the countributions of $I_{ct}$ have to be included, and computed in the full $d+2$ dimensional theory as $\zeta \rightarrow 0$. However, after renormalizing to the boundary at $\zeta= \zeta_b$, no remaining contributions from $I_{ct}$ are needed. Also, such counterterms are not needed for the action in Eq.~(\ref{2d1}) to yield consistent local equations of motion.

We already know from Section~\ref{sec:ads2} that the saddle point metric in the near-horizon region is AdS$_2$. In the form in Eq.~(\ref{dsds2}), the two-dimensional metric is scaled by factor $[\Phi (\zeta \rightarrow \infty)]^{d-1} = R_h^{d-1}$ from that in Eq.~(\ref{ads2}):
\beq
ds_2^2 = \frac{R_2^2 R_h^{d-1}}{\zeta^2} \left[ (1 - 4 \pi^2 T^2 \zeta^2) d\tau^2 + \frac{d \zeta^2}{1 - 4 \pi^2 T^2 \zeta^2} \right] \,. \label{ads2a}
\eeq
Also in the near-horizon regions (A) and (B), the field coupling to $\mathcal{R}_2$, which in our case is $\Phi^d$, has the saddle point obtained from Eq.~(\ref{Phizeta})
\beq
[\Phi (\zeta)]^d =  \Phi_0 + \frac{\Phi_1}{\zeta} + \ldots \quad, \quad R_h \ll \zeta < \frac{1}{2 \pi T}\,, \label{MSY}
\eeq
with the coefficients
\beq
\Phi_0 = R_h^d \quad, \quad \Phi_1 = d R_h^{d-1} R_2^2 \,.
\label{Phib}
\eeq
Note that $\Phi_1/\zeta \ll \Phi_0$ in the entire AdS$_2$ region, including both its bulk (A) and its boundary (B), as in Fig.~\ref{fig:bh}. The solution with $\Phi^d = \Phi_0$ and the metric in Eq.~(\ref{ads2a}) describes the near-horizon AdS$_2 \times S_d$ region (A) in Fig.~\ref{fig:bh}, and we are interested in the structure of the corrections from $\Phi_1$ to this leading order result.

One of the remarkable observations of Almheiri and Polchinski \cite{AAJP15} and Maldacena {\it et al.\/} \cite{JMDS16b} is that the action for quantum fluctuations with these corrections, with general $U(\Phi)$, is universal. More specifically, they argued that the field coupling to $\mathcal{R}_2$ must have the saddle point spatial dependence in Eq.~(\ref{MSY}), and that the action for the quantum fluctuations reduces to a boundary action in region (B) dependent only upon the value of $\Phi_1$ in Eq.~(\ref{MSY}). The independence of the bulk action in region (A) on the $\Phi_1/\zeta$ correction in Eq.~(\ref{MSY}) follows from the first order variation in the action $I_{EM}$ in Eq.~(\ref{2d1}), which vanishes because of the bulk equation of motion for $\Phi_0$
\beq
\delta I_{EM} =  \left[- \frac{s_d}{2\kappa^2} \, d\Phi_0^{d-1} \, \mathcal{R}_2 + U'(\Phi_0) + \frac{Z'(\Phi_0)}{4 g_F^2} F^2\right] \frac{\Phi_1}{\zeta} = 0 \,. \label{Phieom}
\eeq
The result in Eq.~(\ref{Phieom}) is easily verified after employing the AdS$_2$ metric in Eq.~(\ref{ads2a}), the near-horizon gauge field in Eq.~(\ref{Azeta}), and potentials in Eq.~(\ref{2d2}).

Another important observation of Maldacena {\it et al.\/} \cite{JMDS16b} is 
that quantum fluctuations about the metric in Eq.~(\ref{ads2a}) can be represented entirely by fluctuations of a quantum boundary theory (such as the complex SYK model). In the bulk inside the boundary, the metric remains fixed at that in Eq.~(\ref{ads2a}), and the induced metric on the boundary is fixed at $R_2^2 R_h^{d-1}/\zeta_b^2$. The fluctuations of the boundary theory are realized by a boundary time diffeomorphism, which also determines the shape of the boundary embedded in AdS$_2$.
Before determining the action for such fluctuations, we change notation for the bulk time from $\tau$ to $f$, 
and use $\tau$ as the symbol for the parametric time along the boundary. Then the boundary curve is at bulk co-ordinates $(f(\tau), \zeta(\tau))$. The boundary metric induced by Eq.~(\ref{ads2a}) equals $R_2^2 R_h^{d-1}/\zeta_b^2$ after 
we choose, in an expansion in $\zeta_b$,
\beq
\zeta (\tau) = \zeta_b f' (\tau) + \zeta_b^3 \left( \frac{\left[f'' (\tau) \right]^2}{2 f'(\tau)} - 2 \pi^2 T^2  \left[f' (\tau) \right]^3 \right) + \ldots\,. \label{zetacurve}
\eeq
Finally, we evaluate $I_{GH}$ in Eq.~(\ref{2d1}) along this boundary curve. As we have already included the contribution of $\Phi_0$ in Eq.~(\ref{MSY}) at the saddle point, and so we need only include  $\Phi^d \rightarrow \Phi_1/\zeta_b$ in Eq.~(\ref{2d1}). In this manner we obtain the action \cite{JMDS16b} (see Appendix~\ref{app})
\bea
I_{1,{\rm eff}} [f] &=& -  \frac{s_d \Phi_1}{\kappa^2} \int_0^{1/T} d \tau \, \left( \{ f(\tau), \tau\}
+ 2 \pi^2 T^2 \left[f' (\tau) \right]^2 \right)
\nn
&=&   -  \frac{s_d \Phi_1}{\kappa^2} \int_0^{1/T} d \tau \, \{ \tan (\pi T f(\tau)), \tau\}\,. \label{Seff1}
\eea
Note that the function $\tan (\pi T f(\tau))$ in the last equation is the same as that obtained in Eq.~(\ref{map2}) in the co-ordinate mapping from $T=0$ to $T>0$ near the boundary.
Comparing with the action in Eq.~(\ref{Seff}), we obtain 
\beq
\gamma = \frac{4 \pi^2 s_d \Phi_1}{\kappa^2}\,. \label{gammares2}
\eeq
After using the value of $\Phi_1$ in Eq.~(\ref{Phib}), we find that this value of $\gamma$ is in perfect agreement with the value obtained from the thermodynamics of the Einstein-Maxwell theory in $d+2$ dimensions, which is presented in
Eqs. (\ref{gammares1}) and (\ref{gammares}). This is the main result of this section.

We note here that upon evaluating the Schwarzian for $f(\tau) = \tau$, we obtain $I_{1,{\rm eff}} = - \gamma T/2$, which yields a change in the free energy $F = \Omega + \mu \mathcal{Q}$ of
\beq
\Delta F = -\gamma T^2/2. \label{DeltaF}
\eeq
We are working here at constant $\mathcal{Q}$, and hence $I_{1,{\rm eff}}$ contributes to $F$, and not directly to $\Omega$. This result for $\Delta F$ was indicated in Fig.~\ref{fig:bh}, and its $T$-derivative is in Eq.~(\ref{defgamma}).

\section{Effective action for the phase mode}
\label{sec:phase}

This section will consider gauge fluctuations of the Einstein-Maxwell action in Eq.~(\ref{EM}).
These correspond to charge fluctuations in the boundary theory, which are represented by a phase field $\phi$. As in Section~\ref{sec:2dgravity}, we will limit our consideration to $d=2$ in the present section.

We are interested in bulk solutions satisfying the boundary condition 
\beq
A_\tau (\tau, r \rightarrow \infty) = i\mu (\tau) \,,  \label{amu}
\eeq
which is satisfied by Eq.~(\ref{a1}). It is useful to consider the more general case in which $\mu$ is time-dependent, as indicated in Eq.~(\ref{amu}); but we will ultimately make $\mu$ time independent. The key observation of Son and Nickel \cite{Nickel:2010pr} (see also Ref.~\cite{Moitra:2018jqs}) is that there are a family of bulk gauge fields satisfying these boundary conditions. In particular there is a non-trivial Wilson line from the horizon to the boundary which defines the phase field $\phi$ in Eq.~(\ref{Seff}) with non-trivial dynamics
\begin{equation}
\phi(\tau)= \int _{r_0}^{\infty} dr  A_r(\tau,r) \label{defphi}
\end{equation}
Gauge transformations which maintain Eq.~(\ref{amu}) only perform a time-independent shift $\phi(\tau) \rightarrow \phi(\tau) + \mbox{constant}$, corresponding to the presence of a globally conserved U(1) on the boundary. So the effective action of for $\phi$ will depend only on $\partial_\tau \phi$, as we will see below. 
\paragraph{} To derive an effective action for this mode, let us introduce the bulk analogue of this Wilson line
\begin{equation}
\Phi_1 (\tau, r)= \int _{r_0}^{r} dr  A_r(\tau,r)  \;\;\;\;,\;\;\;\;  A_r(\tau, r) = \partial_{r} \Phi_1(\tau, r)
\end{equation}
so that
\beq 
F_{r \tau} = \partial_r \left(A_\tau - \partial_\tau \Phi_1 \right)\,.
\eeq
The bulk field $\Phi_1 (\tau, r)$ acts as a proxy for the radial gauge field and approaches the boundary Wilson line as $r \rightarrow \infty$ {\it i.e.}
\beq
\phi(\tau) \equiv \Phi_1 (\tau, r \rightarrow \infty) \,. 
\eeq
In the presence of a time-dependent $\Phi_1$, we write the metric in Eqs.~(\ref{s1}) as
\beq
ds^2 =  g_{\tau\tau}  d\tau^2 +   g_{rr} dr^2  + r^2 d \Omega_d^2  \,, \label{s1a}
\eeq
where, for now, we allow for arbitrary $\tau$ and $r$ dependence in $g_{\tau\tau}$ and $g_{rr}$.
Then the Maxwell term in the action in Eq.~(\ref{EM}) can be written as
\beq
I_M = \frac{s_d}{2g_F^2} \int dr d \tau \frac{r^d}{\sqrt{g_{\tau\tau} g_{rr}}} F_{r \tau}^2 \,
\eeq
where we have integrated over the angular co-ordinates.
Now let us examine the bulk equation of motion for $A_\tau$
\begin{equation}
\partial_{r} \left( \frac{r^d}{\sqrt{g_{\tau\tau} g_{rr}}} F_{r\tau} \right) = 0 \;\;\; \implies \;\;\;\; F_{r\tau} = c_1 (\tau) \frac{\sqrt{g_{\tau\tau} g_{rr}}}{r^d}
\end{equation}
We can determine the function $c_1 (\tau)$ by integrating the second equation to obtain
\beq
i \mu - \partial_\tau \phi = c_1 (\tau) \int_{r_0}^{\infty} dr \frac{\sqrt{g_{\tau\tau} g_{rr}}}{r^d}\,. \label{valc1}
\eeq
This determines $c_1 (\tau)$ in terms of the metric and the combination $i \mu - \partial_\tau \phi$. We can insert the $c_1 (\tau)$ so determined into the Maxwell action and obtain
\beq
I_M = \frac{s_d}{2g_F^2} \int dr d \tau
\left[c_1 (\tau)\right]^2 \frac{\sqrt{g_{\tau\tau} g_{rr}}}{r^d}
\label{nIM}
\eeq

We now need to insert the Maxwell action specified by Eqs.~(\ref{valc1}) and (\ref{nIM}) into Eq.~(\ref{EM}), and solve the resulting saddle point equations for the metric obtained from the total action $I_{EM} + I_{GH} + I_{ct}$. At zeroth order in $\partial_\tau \phi$, this solution is just that specified by Eq.~(\ref{s1}). However, we need to determine the correction to the action to order $(\partial_\tau \phi)^2$, and for this we need to include the corrections to the metric which are linear order in $\partial_\tau \phi$; in the boundary theory, these corrections correspond to perturbations in the stress energy tensor which are sourced by a non-zero $\partial_\tau \phi$. Fortunately, these corrections, and the resulting change in the effective action, can be determined by a simple argument. Notice that the influence of $\partial_\tau \phi$ is solely by the shift $\mu \rightarrow \mu + i \partial_\tau \phi$ in Eq.~(\ref{valc1}). At low frequencies, it is safe to ignore the time-dependence in $\partial_\tau \phi$, and so the shift in the 
metric is simply proportional to the $\mu$ derivative of the metric (which is non-zero). 
So we can compute the action by working at a fixed $\mu$, and then replacing $\mu \rightarrow \mu + i \partial_\tau \phi$.

The combined contribution to the effective action from $\partial_\tau \phi$ fluctuations at $T=0$ is then
\bea
I_{2,{\rm eff}} &=& \int d \tau \left[ \Omega (\mu_0 + i \partial_\tau \phi, T=0) - \Omega (\mu_0, T=0) \right] \nn
&=& \int d \tau \left[ - i \mathcal{Q} \, \partial_\tau \phi  + \frac{K}{2} \left( \partial_\tau \phi \right)^2  + \ldots \right]\, \label{I2eff}
\eea
where $\Omega$ is the grand potential in Eq.~(\ref{Omega}).
The second term is a total derivative, and the last term has a coefficient which equals the compressibility $K$ in Eq.~(\ref{defK}).

\subsection{Non-zero temperatures}
\label{sec:phaseT}
This section describes the extension of the phase action in Eq.~(\ref{I2eff}) to $T>0$.

At $T=0$, we have imposed the rigid boundary condition $\mu = \mu_0$ in all our analysis so far, and this fixes the form of $I_{2,\rm eff}$ to that in Eq.~(\ref{I2eff}).
The situation changes at $T >0$, because the computations in Sections~\ref{sec:ads2} and \ref{sec:2dgravity} assumed a fixed $\mathcal{Q}$ and a variable $T>0$, and this required the $T$-dependent change in chemical potential in Eq.~(\ref{muET}). In contrast, in Section~\ref{sec:phase} so far, we are considering the effective action at fixed $\mu$ and variable $T$. In terms of boundary conditions in the AdS/CFT context, these situations correspond to whether we fix the co-efficient of $r^0$ term in Eq.~(\ref{a1}) (as in Section~\ref{sec:phase}) or the co-efficient of the $r^{1-d}$ term (as in Sections~\ref{sec:ads2} and \ref{sec:2dgravity}) as we vary $T$.

Therefore, we need to supplement the fixed $\mathcal{Q}$ action in Eq.~(\ref{Seff1}), with a fixed $\mu$ action. It is useful to motivate the required action by considering the relationship between the corresponding thermodynamic derivatives. We saw in Eq.~(\ref{DeltaF}) that the Schwarzian action computed $(\partial^2 F/\partial T^2)_{\mathcal{Q}}$. Correspondingly, we wish to extend Eq.~(\ref{I2eff}) to compute $(\partial^2 \Omega/\partial T^2)_{\mathcal{\mu}}$. But the difference between these two derivatives is specified by thermodynamics:
\bea
\left( \frac{\partial^2 \Omega}{\partial T^2} \right)_\mu &=&
\left( \frac{\partial^2 F}{\partial T^2} \right)_\mathcal{Q} + \left( \frac{\partial^2 \Omega}{\partial \mu^2} \right)_T \left[ \left( \frac{\partial \mu}{\partial T} \right)_{\mathcal{Q}} \right]^2 \nn
&=& \left( \frac{\partial^2 F}{\partial T^2} \right)_\mathcal{Q}  - K \left[ \left( \frac{\partial \mu}{\partial T} \right)_{\mathcal{Q}} \right]^2 \quad , \quad T \rightarrow 0\,. \label{thermo}
\eea
We can now assume that both free energies are time integrals of their respective actions, and as above Eq.~(\ref{I2eff}), we momentarily ignore the frequency dependence of the actions. Then 
we can carry out the mapping in Eq.~(\ref{thermo}) between the fixed $\mathcal{Q}$ and fixed $\mu$ situations at the level of local effective actions. At order $T^2$, such an analysis amounts to replacing $({\partial \mu}/{\partial T})_{\mathcal{Q}}$ by the difference in the chemical potential between the two approaches divided by $T$. 
So we need to take the difference between the chemical potential in the fixed $\mu$ case, {\it i.e.\/} $\mu_0 + i \partial_\tau \phi$, from that in the fixed $\mathcal{Q}$ case, {\it i.e.\/}  $\mu_0 - 2 \pi \mathcal{E} T$ as in Eq.~(\ref{muET}). Their difference is $i \partial_\tau \phi + 2 \pi \mathcal{E} T$, and this identifies the required modification of $I_{2,{\rm eff}}$:
\beq 
I_{3,{\rm eff}} = \frac{K}{2} \int d \tau  \left( \partial_\tau \phi - i2 \pi \mathcal{E} T \right)^2 \,. \label{I3eff}
\eeq
Now at leading order, ignoring phase fluctuations, we obtain a contribution from Eq.~(\ref{I3eff}) to the grand potential of $- 2 \pi^2 K \mathcal{E}^2 T^2$. By Eq.~(\ref{thermo}) this has to be added to the contribution in Eq.~(\ref{DeltaF}), to yield the total order $T^2$ contribution to the grand potential 
\beq
\Delta\Omega = - (\gamma + 4 \pi^2 K \mathcal{E}^2)  \frac{T^2}{2} \,. \label{DeltaOmega}
\eeq
It can now be verified that Eqs.~(\ref{DeltaOmega}) and (\ref{DeltaF}) are consistent with Eq.~(\ref{thermo}), and also with the explicit value of $(\partial^2 \Omega/\partial T^2)_\mu$ in Eq.~(\ref{DeltaOmega0}), after using the values of $\gamma$, $K$, and $\mathcal{E}$ in Section~\ref{sec:bh}.

\subsection{Coupling to the diffeomorphism mode}
\label{sec:diffeo}

This section considers the modification of the phase action in Eq.~(\ref{I3eff}) from the boundary time diffeomorphism mode of Section~\ref{sec:2dgravity}. 

An important observation, following from the analysis above Eq.~(\ref{I3eff}), 
is that any coupling of Eq.~(\ref{I3eff}) to a diffeomorphism mode should vanish at $T=0$.
There can be no corrections to Eq.~(\ref{I2eff}) at $T=0$, apart from a renormalization of the coupling $K$, and the effective action can only depend upon the combination $\mu + i \partial_\tau \phi$ independent of the metric. 

In Section~\ref{sec:phaseT}, we argued that we need the chemical potential which keeps $\mathcal{Q}$ fixed at variable $T$ for the computation in Section~\ref{sec:2dgravity}. In the absence of diffeomorphisms in time, this was given by Eq.~(\ref{muET}). We will now compute the correction to Eq.~(\ref{muET}) in the presence of the time diffeomorphism of the boundary theory.

For this computation, we focus on the AdS$_2$ region (A) of Fig.~\ref{fig:bh}. The vector potential is given by Eq.~(\ref{Azeta}), which we write as
\beq
A_\tau = -i 2 \pi \mathcal{E} T + i\frac{\mathcal{E}}{\zeta} \,.  \label{Azeta2}
\eeq
We apply the usual rules of the AdS/CFT correspondence \cite{Hartnoll:2016apf} at the AdS$_2$ boundary, which is $\zeta \rightarrow 0$ here (but with $\zeta \gg R_h$, see Fig~\ref{fig:bh}).
We identify the $\zeta^{0}$ term in Eq.~(\ref{Azeta2}) with the chemical potential at the AdS$_2$ boundary, while the coefficient of the $\zeta^{-1}$ is proportional to the conjugate charge density. Notice also that 
the form in Eq.~(\ref{Azeta2}) is asymptotically consistent with the form in Eq.~(\ref{a1}) for AdS$_{d+2}$ with $d=0$ (after mapping from $r$ to $\zeta$ via Eq.~(\ref{rzeta})).

Now let us consider quantum fluctuations on the boundary theory (realized by the complex SYK model), 
represented by the boundary time
diffeomorphism $f(\tau)$. The chemical potential of the boundary, $-iA_\tau^b$, transforms like the time component of a vector potential: 
 $A_\tau^b d\tau = A_f^b df$, and so $A_\tau^b = A_f^b (\partial_\tau f)$. The boundary chemical potential in the original time is $-i A_f^b =  - 2 \pi \mathcal{E} T $, and so the chemical potential in the theory with time $\tau$ is $-i A_\tau^b= -(2 \pi \mathcal{E} T) \partial_\tau f$.

Having computed the chemical potential at the AdS$_2$ boundary, we need to determine the chemical potential at the AdS$_{d+2}$ boundary, $r \rightarrow \infty$. In the absence of time diffeomorphisms, these two chemical potentials are connected by Eq.~(\ref{muET}). We have already argued that there can be no $T$-independent coupling to the diffeomorphism $f(\tau)$. Furthermore, leading $T$-dependent renormalization to $\mu$ in Eq.~(\ref{muET}) arises entirely from the AdS$_2$ region. So we conclude that the generalization of Eq.~(\ref{muET}) is
\beq
\mu = \mu_0 - (2 \pi \mathcal{E} T) \partial_\tau f \quad , \quad \mbox{$T \rightarrow 0$, fixed $\mathcal{Q}$.}
\eeq
Using this renormalized chemical potential in the reasoning above Eq.~(\ref{I3eff}), we obtain the updated action for phase fluctuations
\beq 
I_{4,{\rm eff}} = \frac{K}{2} \int d \tau  \left( \partial_\tau \phi - i (2 \pi \mathcal{E} T) \partial_\tau f \right)^2 \,. \label{I4eff}
\eeq
The full action is therefore $I_{1,{\rm eff}} + I_{4,{\rm eff}}$, which yields Eq.~(\ref{Seff}) from Eq.~(\ref{Seff1}).

\section{Discussion}

The Reissner-N\"ordstrom-AdS charged black hole has been extensively used as a holographic model of strongly interacting quantum matter at non-zero density \cite{Hartnoll:2016apf}. Near the boundary, the geometry is AdS$_{D}$, and so the conventional rules of the AdS/CFT correspondence apply, and they can be used to relate bulk properties to the correlations of the boundary quantum theory in $D-1$ spacetime dimensions. It was also recognized \cite{Faulkner09} that (for $D>3$) the low temperature correlations are linked to the near-horizon AdS$_2$ geometry. But it had not seemed possible to express the physics in terms of the 2-dimensional bulk alone, without embedding it in a higher-dimensional geometry.

Maldacena {\it et al.\/} \cite{JMDS16b} recently proposed a novel formulation of the 2-dimensional bulk quantum physics. Following the example of the SYK model, they argued that the strong back reaction of the AdS$_2$ geometry to external perturbations \cite{AAJP15,Alm2016fws} could be accounted for by integrating over a time diffeomorphism ($f(\tau)$) in the quantum theory on the boundary of AdS$_2$. After fixing the induced metric on the boundary of AdS$_2$, the time diffeomorphism determines the shape of the AdS$_2$ boundary. They also obtained a 0+1 dimensional Schwarzian action for the time diffeomorphisms. For the case of a charged black hole, the bulk U(1) gauge field implies that the Schwarzian action has to be supplemented \cite{Davison17} by that of a scalar phase field ($\phi (\tau)$), leading to the action in Eq.~(\ref{Seff}). The path integral over this action can be exactly computed \cite{Stanford:2017thb,GKST19,Liu:2019niv}, and this allows computation of quantum properties beyond what has been 
possible from the AdS$_D$ approach above.

These advances result in two approaches to determining the low temperature correlations of the quantum system holographically equivalent to a charged black hole: we can use the conventional AdS/CFT correspondence at the AdS$_D$ boundary, or the Schwarzian theory at the AdS$_2$ boundary (see Fig.~\ref{fig:geometry}).
Earlier works \cite{SS10,kitaev2015talk,SS15,JMDS16b,KJ16,HV16,Davison17,Moitra:2018jqs}
established the equivalence of the two approaches by comparing thermodynamics and correlation functions. Here, we have derived the effective 0+1 dimensional action as a low energy limit of the Einstein-Maxwell theory of charged black holes in asymptotically AdS$_{D}$ space, and confirmed that the tree-level predictions of the two actions are in precise quantitative agreement. The quantum fluctuation corrections from the 0+1 dimensional effective action can now be applied to the $D$-dimensional Einstein-Maxwell theory.
The mapping to the effective theory is valid at temperatures $T \ll 1/R_h$, where $R_h$ is the radius of the black hole. However, we do not assume any particular relation between $R_h$ and the AdS$_{D}$ radius $L$, and our analysis can approach asymptotically Minkowski space for large $L/R_h$.

\subsection*{Acknowledgements}
This analysis was undertaken for lectures at the 36th Advanced School in Physics at the Israel Institute for Advanced Studies in Jerusalem (\href{https://www.youtube.com/playlist?list=PLTn74Qx5mPsQHL3o6fC3vCKZ-YaLUvt9O}{lecture videos}), and I am grateful to all participants for many stimulating interactions. 
I thank M.~Blake, R.~Davison, N.~Iqbal, J.~Maldacena, G.~Mandal, D.~Stanford, S.~P.~Trivedi, and S.R.~Wadia for useful discussions. M.~Blake and R.~Davison contributed to the early stages of the analysis in Section~\ref{sec:phase}.
This research was supported by the US Department of Energy under Grant No. DE-SC0019030. Research at Perimeter Institute is supported by the Government of Canada through Industry Canada and by the Province of Ontario through the Ministry of Research and Innovation. I also acknowledges support from Cenovus Energy at Perimeter Institute.

\appendix

\section{Extrinsic curvature and the Schwarzian}
\label{app}

We consider a general metric of a two-dimensional space with co-ordinates $(f,  \zeta)$
\beq
ds^2 = h_f (\zeta) d f^2 +  h_\zeta (\zeta) d\zeta^2\,.
\eeq
We are interested in the extrinsic curvature of a curve $\mathcal{C}$ parameterized by $\tau$: $(f (\tau), \zeta(\tau))$.

Let us transform to new co-ordinates $(\tau, \lambda)$ so that the curve $\mathcal{C}$ is at $\lambda =0$. For small $\lambda$ we choose the co-ordinate transformation
\bea 
f &=& f(\tau) + \lambda  \nn
\zeta &=& \zeta(\tau) - \lambda \frac{h_f (\zeta(\tau)) f'(\tau)}{h_\zeta (\zeta (\tau)) \zeta' (\tau)} + \mathcal{O} (\lambda^2)\,.
\eea
This insures that the metric in the new co-ordinates is of the Gaussian normal form
\beq
ds^2 = h_\lambda (\tau, \lambda) d \lambda^2 + h_b (\tau, \lambda) d \tau^2
\eeq
with
\bea
h_\lambda (\tau, \lambda) &=&  h_f (\zeta(\tau)) + \mathcal{O}(\lambda) \nn
h_b (\tau, \lambda) &=& h_f (\zeta (\tau)) [f'(\tau)]^2 +  h_\zeta (\zeta (\tau)) [\zeta'(\tau)]^2 - 2 \lambda h_\zeta (\zeta(\tau)) 
\zeta'(\tau) \frac{d}{d\tau} \left( \frac{h_f (\zeta(\tau)) f'(\tau)}{h_\zeta (\zeta (\tau)) \zeta' (\tau)} \right) \nn
&~&  - \lambda \frac{h_f (\zeta(\tau)) f'(\tau)\{[f'(\tau)]^2 h_f' (\zeta(\tau)) + [\zeta'(\tau)]^2 h_\zeta' (\zeta(\tau))\}}{h_\zeta (\zeta(\tau)) \zeta' (\tau)} + \mathcal{O}(\lambda^2) \,.
\eea
The induced metric on $\mathcal{C}$ is $h_b (\tau, 0)$, and the extrinsic curvature of $\mathcal{C}$ is
\beq
\mathcal{K}_1 = \frac{1}{2 h_b (\tau, 0) \sqrt{h_\lambda (\tau, 0)}} \left. \frac{d h_b (\tau, \lambda)}{d \lambda} \right|_{\lambda =0}\,.
\eeq
From Eq.~(\ref{ads2a}), we now use
\beq
h_f (\zeta) = \frac{R_2^2 R_h^{d-1}}{\zeta^2} (1 - 4 \pi^2 T^2 \zeta^2) \quad , \quad  h_\zeta (\zeta) = \frac{R_2^2 R_h^{d-1}}{\zeta^2} \frac{1}{(1 - 4 \pi^2 T^2 \zeta^2)} \,,
\eeq
and fix the curve $\mathcal{C}$ by Eq.~(\ref{zetacurve}) which sets $h_b (\tau, 0) = R_2^2 R_h^{d-1}/\zeta_b^2$. Then we evaluate the extrinsic curvature of $\mathcal{C}$, expand in powers of $\zeta_b$, and insert in Eq.~(\ref{2d1}) with $\Phi^d \rightarrow \Phi_1/\zeta_b$, to obtain Eq.~(\ref{Seff1}). Note ({\it i\/}) all the powers of the metric prefactor $R_2^2 R_h^{d-1}$ cancel out;
({\it ii\/}) in the evaluation of $\mathcal{K}_1$ (but not for other quantities), it turns out we only need to keep the leading term of order $\zeta_b^1$ in Eq.~(\ref{zetacurve}).

\bibliographystyle{JHEP}

\bibliography{syk}

\end{document}